\documentclass[12pt]{article}

\usepackage{amssymb}
\usepackage{amsmath}
\usepackage{amscd}
\usepackage{latexsym}
\usepackage{graphicx}

\usepackage{cite}

\newcommand{\be}{\begin{equation}}
\newcommand{\ee}{\end{equation}}

\newcommand{\dlt}{\delta}
\newcommand{\prt}{\partial}
\newcommand{\br}{{\bf r}}

\newcommand{\bq}{{\bf q}}
\newcommand{\bp}{{\bf p}}
\newcommand{\ba}{{\bf a}}
\newcommand{\bt}{\beta}
\newcommand{\vp}{\varphi}
\newcommand{\ep}{\varepsilon}
\newcommand{\al}{\alpha}
\newcommand{\ra}{\rightarrow}

\newcommand{\gm}{\gamma}

\newcommand{\Om}{\Omega}

\newcommand{\dgr}{\dagger}

\newcommand{\cD}{{\cal D}}

\newcommand{\cH}{{\cal H}}

\newcommand{\cA}{{\cal A}}

\newcommand{\rgl}{\rangle}
\newcommand{\lgl}{\langle}

\begin{document}

\begin{center}

{\Large{\bf Statistical theory of structures with extended defects} \\ [5mm]

V.I. Yukalov$^{1,2}$ and E.P. Yukalova$^{3}$ }  \\ [3mm]

{\it 
$^1$Bogolubov Laboratory of Theoretical Physics, \\
Joint Institute for Nuclear Research, Dubna 141980, Russia \\ [2mm]

$^2$Instituto de Fisica de S\~ao Carlos, Universidade de S\~ao Paulo, \\
CP 369,  S\~ao Carlos 13560-970, S\~ao Paulo, Brazil \\ [2mm]

$^3$Laboratory of Information Technologies, \\
Joint Institute for Nuclear Research, Dubna 141980, Russia } \\ [3mm]

{\bf E-mails}: {\it yukalov@theor.jinr.ru}, ~~ {\it yukalova@theor.jinr.ru}

\end{center}

\vskip 1cm

\begin{abstract}
Many materials contain extended defects of nanosize scale, such as dislocations, 
cracks, pores, polymorphic inclusions, and other embryos of competing phases. 
When one is interested not in the precise internal structure of a sample with 
such defects, but in its overall properties as a whole, one needs a statistical 
picture giving a spatially averaged description. In this chapter, an approach 
is presented for a statistical description of materials with extended nanosize 
defects. A method is developed allowing for the reduction of the problem to the 
consideration of a set of system replicas representing homogeneous materials 
characterized by effective renormalized Hamiltonians. This is achieved by 
defining a procedure of averaging over heterophase configurations. The method 
is illustrated by a lattice model with randomly distributed regions of disorder.     
\end{abstract}

\newpage

\section{Introduction}

The structure of many solid-state materials is not represented by ideal 
crystalline lattices, but contains various defects 
\cite{Schulze_1,Ziman_2,Filshtinsky_3}. By 
their size, defects can be classified into two types, point defects and 
extended defects. Examples of point defects are Schottky defects, Frenkel 
defects, vacancies, interstitials, and impurities. Examples of extended 
defects are dislocations, cracks, pores, heterophase embryos, and polymorphic 
inclusions. These defects are usually of nanosize scale, at least in some 
directions. For instance, a dislocation length can be of a macroscopic size, 
while a dislocation radius is of a nanoscale size \cite{Hirth_4,Hull_5}.

In many cases, one is interested in the overall properties of a sample with 
nanosize defects, but not in the details of its internal structure. In that 
case, one needs a theory describing the sample as a whole. This implies that 
it is necessary to have an approach characterizing the average typical features 
of the sample. That is, there is a need in a kind of a statistical approach. 

In this chapter, we present such a statistical approach for describing 
materials with nanosize defects. We keep in mind materials whose ideal 
crystalline structure is disturbed by the presence of extended defects with 
a disordered structure. These regions of disorder are randomly distributed 
across the sample. Usually, they are in local equilibrium or in quasi-equilibrium 
with the surrounding crystalline matrix. In some cases they can 
move through the sample, as e.g. dislocations through a crystal. The fraction 
of particles forming the disordered regions, with respect to the total number 
of particles in the system, generally, is defined self-consistently from the 
conditions of the material stability.   

Throughout the chapter, the system of units is used where the Planck and 
Boltzmann constants are set to one.

\section{Spatial separation of phases}

The spatial regions filled by extended defects can be treated as the regions 
filled by a phase with disordered particle locations. We consider a sample 
containing a mixture of two different phases composed of the same kind of 
particles. Say, one phase, forms a solid with a crystalline lattice, and 
the other phase consists of regions of a disordered structure. Schematically, 
this is shown in Fig. 1. Using the Gibbs method of separating surfaces 
\cite{Gibbs_6}, the whole sample can be represented as consisting of the 
regions filled by the corresponding phases, so that the total sample volume 
is the sum of the volumes filled by each of the phases and the total number 
of particles is the sum of the particles in each phase,
\be
\label{1}
 V = V_1 + V_2 \; , \qquad N = N_1 + N_2 \; .
\ee
The regions of disorder are randomly distributed in space.

Since we are interested in the average properties of the system, hence we need 
to describe the related averaging procedure. Below we present such a procedure 
based on the review articles \cite{Yukalov_7,Yukalov_8}. For generality, we 
consider a quantum system. The same approach can also be realized for classical 
systems.

The spatial location of the phases can be characterized by the manifold indicator 
functions \cite{Bourbaki_9}
\begin{eqnarray}
\label{2}
\xi_f(\br) = \left\{ \begin{array}{ll}
1 , ~ & ~ \br \in V_f \\
0 , ~ & ~ \br \not\in V_f \; .
\end{array} \right.
\end{eqnarray}
Here, for the simplicity of notation, we denote the spatial part, occupied by 
an $f$-th phase, and the volume of this part by the same letter $V_f$, with 
$f=1,2$. 

The Hilbert space of microscopic states is the tensor product
\be
\label{3}
  \cH = \cH_1 \; \bigotimes \; \cH_2 
\ee
of the Hilbert spaces associated with the phases enumerated by the index 
$f=1,2$. The spaces $\cH_f$ can be defined as weighted Hilbert spaces 
\cite{Yukalov_7,Yukalov_10}. The algebra of observables $\cA(\xi)$ can be 
represented as a direct sum of the algebras of observables $\cA_f(\xi_f)$ 
on the spaces $\cH_f$ for the related phases,
\be
\label{4}
 \cA(\xi) = \cA_1(\xi_1) \; \bigoplus \; \cA_2(\xi_2) \; ,
\ee
where $\cA_f(\xi_f)$ is the representation of the algebra on the Hilbert 
space $\cH_f$. Respectively, the energy operator (Hamiltonian) has the form
\be
\label{5}
\hat H(\xi) = \hat H_1(\xi_1) \; \bigoplus \; \hat H_2(\xi_2) \; .
\ee
And the number-of-particle operator is
\be
\label{6}
\hat N(\xi) = \hat N_1(\xi_1) \; \bigoplus \; \hat N_2(\xi_2) \; .
\ee
 
The general expression of the energy operator for an $f$-th phase reads as
$$
\hat H_f(\xi_f) = \int \xi_f(\br) \; \psi_f^\dgr(\br) \left[ -\; 
\frac{\nabla^2}{2m} + U(\br) \right] \; \psi_f(\br) \; d\br \; + 
$$
\be
\label{7}
+ \; \frac{1}{2} \int \xi_f(\br) \xi_f(\br') \; 
\psi_f^\dgr(\br) \psi_f^\dgr(\br') \; \Phi(\br-\br') \; 
\psi_f(\br') \psi_f(\br) \; d\br d\br' 
\ee
and the number-of-particle operator for an $f$-th phase has the form
\be
\label{8}
\hat N_f(\xi_f) = 
\int \xi_f(\br) \; \psi_f^\dgr(\br) \psi_f(\br) \; d\br \; .
\ee
Here $U({\bf r})$ is an external potential and $\Phi({\bf r})$ is an interaction 
potential.

\section{Statistical operator of mixture}

The statistical operator $\hat{\rho}(\xi)$ for a heterophase system depends on 
the configuration of the phases in the sample, which is denoted by $\xi$. This 
operator can be found from the principle of minimal information by minimizing 
the information functional under given additional constraints. These are: the 
normalization condition
\be
\label{9}
{\rm Tr} \int \hat\rho(\xi)\; \cD(\xi) = 1 \; ,
\ee
the definition of the system energy
\be
\label{10}
 {\rm Tr} \int \hat\rho(\xi) \hat H(\xi) \; \cD(\xi) = E \; ,
\ee
and the total number of particles
\be
\label{11}
 {\rm Tr} \int \hat\rho(\xi) \hat N(\xi) \; \cD(\xi) = N \; .
\ee
Here the trace is over the total Hilbert space (\ref{3}). The notation 
$\cD(\xi)$ implies the averaging over all admissible phase configurations that 
should be done in view of the random locations and shapes of the co-existing 
phases.   
   
The corresponding information functional can be taken in the Kullback-Leibler 
\cite{Kullback_11,Kullback_12} form
$$
I[\; \hat\rho\;] = {\rm Tr} \int \hat\rho(\xi) \; 
\ln \; \frac{\hat\rho(\xi)}{\hat\rho_0(\xi)} \; \cD(\xi) \; + \;
\al \left[  {\rm Tr} \int \hat\rho(\xi) \; \cD(\xi) \; - \; 1\right] \; + 
$$
\be
\label{12}
+ \; 
\bt \left[  {\rm Tr} \int \hat\rho(\xi) \hat H(\xi)\; \cD(\xi) \; - 
\; E \right] \; +  \; 
\gm \left[  {\rm Tr} \int \hat\rho(\xi) \hat N(\xi)\; \cD(\xi) \; - 
\; N \right] \; ,
\ee
where $\al$, $\bt\equiv 1/T$, and $\gm\equiv-\bt\mu$ are Lagrange multipliers, 
$T$ is temperature, and $\hat{\rho}_0(\xi)$ is a trial statistical operator 
prescribed by some apriori information, if any. In the case of no preliminary 
information, $\hat{\rho}_0(\xi)$ is a constant. Then the minimization of the 
information functional yields the statistical operator
\be
\label{13}
\hat\rho(\xi) = \frac{1}{Z} \; \exp\{ - \bt H(\xi) \} \; ,
\ee
with the partition function
\be
\label{14}
 Z = {\rm Tr} \int  \exp\{ - \bt H(\xi) \} \; \cD(\xi) \; ,
\ee
where the grand Hamiltonain is 
\be
\label{15}
 H(\xi) \equiv \hat H(\xi) - \mu \hat N(\xi) \;  .
\ee
Defining the effective renormalized Hamiltonian by the relation
\be
\label{16}
\int  \exp\{ - \bt H(\xi) \} \; \cD(\xi) = \exp\{-\bt \widetilde H \}
\ee
results in the partition function
\be
\label{17}
 Z = {\rm Tr}\; \exp\{ - \bt \widetilde H \} \; .
\ee
From here, the grand thermodynamic potential follows:
\be
\label{18}
 \Om = - T \; \ln Z \;  .
\ee

Recall that this thermodynamic potential characterizes a system where the 
regions of disorder are not frozen and their spatial locations are not fixed. 
This implies that the experiment with such a system does not provide information 
on spatial locations of the phase distribution, but the experimental data give 
a spatially averaged picture. This is why it is necessary to average over phase 
configurations, as is discussed above. In that sense, the overall properties of 
the averaged system correspond to an effectively equilibrium system, although 
at each moment of time the heterophase system is quasi-equilibrium. A multiphase 
system equilibrium on average can be called a system that is in {\it heterophase 
equilibrium} \cite{Yukalov_7}.

\section{Quasi-equilibrium snapshot picture}

As is explained above, we consider the case of a heterophase system equilibrium 
on average, where the regions of disorder are randomly distributed over the 
sample and, generally, can move in space. For example, dislocations, strictly 
speaking, are intrinsically nonequilibrium objects that are created, annihilated, 
and moving through the sample 
\cite{Cottrell_14,Friedel_15,Hirth_16,Langer_17,Langer_18,Langer_19}. However 
a system with these dislocations, being averaged over their locations, represents 
an equilibrium crystal. The approach, we have started describing, is appropriate 
for the interpretation of those experiments that give the results averaged over 
random locations of phase configurations. 

It is instructive to compare this approach with the description of a sample 
containing different phases whose locations are fixed in space. From the 
experimental point of view, the latter situation corresponds to the case 
where the observed data are taken in a snapshot way. This is called the case 
of spatially frozen phases. For the latter case, thermodynamics, generally, 
differs from the described above. 

For a system with frozen phases, it is possible to define the statistical 
operator resorting to a quasi-equilibrium picture \cite{Yukalov_7,Yukalov_13}. 
The phase locations can again be described by a set of the manifold indicator 
functions (\ref{2}). With the statistical operator of the whole system 
$\hat{\rho}(\xi)$, the statistical operators for each phase read as
\be
\label{19}
 \hat\rho_f(\xi_f) \equiv {\rm Tr}_{\cH/\cH_f}\; \hat\rho(\xi) \;  .
\ee
This operator has to be normalized,
\be
\label{20}
{\rm Tr}_{\cH_f}\; \hat\rho_f(\xi_f)  = 1 \; .
\ee
For each of the phases, the definitions of the energy
\be
\label{21}
{\rm Tr}_{\cH_f}\; \hat\rho_f(\xi_f) \hat H_f(\xi_f)  = E_f(\xi_f)
\ee
and the number of particles
\be
\label{22}
{\rm Tr}_{\cH_f}\; \hat\rho_f(\xi_f) \hat N_f(\xi_f)  = N_f(\xi_f)   
\ee
are given. 

The information functional takes the form
$$
I[\; \hat\rho_f\;] = {\rm Tr}_{\cH_f}  \hat\rho_f(\xi_f) \; 
\ln \; \frac{\hat\rho_f(\xi_f)}{\hat\rho_f^0(\xi_f)}  \; + \; \al_f(\xi_f) 
\left[  {\rm Tr}_{\cH_f} \hat\rho_f(\xi_f) \;  - \; 1\right] \; + 
$$
\be
\label{23}
 + \; 
\bt_f(\xi_f) \left[  {\rm Tr}_{\cH_f} \hat\rho_f(\xi_f) \hat H_f(\xi_f)\;  
- \; E_f(\xi_f) \right] \; +  \; 
\gm_f(\xi_f) \left[  {\rm Tr}_{\cH_f} \hat\rho_f(\xi_f) \hat N_f(\xi_f)\;  
- \; N_f(\xi_f) \right] \;  ,
\ee
where $\hat{\rho}_f^0(\xi_f)$ is a trial statistical operator. As earlier, we 
assume that no additional information is provided, so that the trial operator is 
constant. Now the Lagrange multipliers depend on the given phase distribution. 
For instance, temperature becomes
\be
\label{24}
 T_f(\xi_f) = \frac{1}{\bt_f(\xi_f)} \; .
\ee
Using the notation
\be
\label{25}
 \gm_f(\xi_f) \equiv - \bt_f(\xi_f)\; \mu_f(\xi_f) \;  ,
\ee
we minimize the information functional, which results in the statistical operator
\be
\label{26}
 \hat\rho_f(\xi_f) = \frac{1}{Z_f(\xi_f)} \; 
\exp\left\{ - \bt_f(\xi_f) H_f(\xi_f) \right\} \;  ,
\ee
with the partition function
\be
\label{27}
 Z_f(\xi_f) = {\rm Tr}_{\cH_f} \; 
\exp\left\{ -\bt_f(\xi_f) H_f(\xi_f) \right\} 
\ee
and the grand Hamiltonian
\be
\label{28}
 H_f(\xi_f) = \hat H_f(\xi_f) - \mu_f(\xi_f) \hat N_f(\xi_f) \; .
\ee

Thus for the frozen phase distribution, for each phase, we get the grand 
thermodynamic potential
\be
\label{29}
 \Om_f(\xi_f) = - T(\xi_f) \ln \; Z_f(\xi_f) \;  .
\ee
Averaging it over phase configurations, we obtain the thermodynamic potential 
for the whole system with the frozen phases,
\be
\label{30}
\overline \Om = \int \; \sum_f \Om_f(\xi_f) \; \cD(\xi) \; .
\ee

As is clear, the grand thermodynamic potentials (\ref{18}) and (\ref{30}), 
in general, do not coincide. However, under some additional restrictions (to 
be mentioned below), the thermodynamic potential (\ref{18}) can approximate 
potential (\ref{30}).

\section{Averaging over phase configurations}

The averaging over phase configurations implies the functional integration 
over the manifold indicator functions (\ref{2}). This integration has been 
explicitly formulated and accomplished in the series of papers 
\cite{Yukalov_7,Yukalov_20,Yukalov_21,Yukalov_22,Yukalov_23,Yukalov_24}. 
Here we summarize the main results of this functional integration. 

First it is necessary to consider the averaging of the functionals of the 
often met form
\be
\label{31}
C_f(\xi_f) = \sum_{m=0}^\infty \int \xi_f(\br_1) \xi_f(\br_2) \ldots
\xi_f(\br_m) \; C_f(\br_1,\br_2,\ldots,\br_m) \; d\br_1  d\br_2 \ldots
 d\br_m \; .
\ee
Then we give the expression for the thermodynamic potential (\ref{18}). And 
finally the averages of the operators corresponding to observable quantities 
are derived. 

\vskip 2mm

{\bf Theorem 1}. The averaging over phase configurations of functional 
(\ref{31}) gives
\be
\label{32}
 \int C_f(\xi_f) \; \cD\xi = C_f(w_f) \; ,
\ee
where
\be
\label{33}
 C_f(w_f) = \sum_{m=0}^\infty \; w_f^m 
\int C_f(\br_1,\br_2,\ldots,\br_m) \; d\br_1  d\br_2 \ldots d\br_m  \; ,
\ee
and the weight
\be
\label{34}
 w_f \equiv \frac{1}{V} \int \xi_f(\br) \; d\br  
\ee
defines the geometric probability of an $f$-th phase. 

\vskip 2mm

{\bf Theorem 2}. The thermodynamic potential
\be
\label{35}
 \Om = - T \ln \; {\rm Tr} \int \exp\{ - \bt H(\xi) \} \; \cD\xi \; ,
\ee
after the averaging over phase configurations, becomes
\be
\label{36}
 \Om = - T \ln \; {\rm Tr} \exp\{ - \bt \widetilde H \}  = 
\sum_f \Om_f \;  ,
\ee
where
\be
\label{37}
\Om_f = - T \ln \; {\rm Tr}_{\cH_f} \exp\{ - \bt H_f(w_f) \} \; .
\ee
The renormalized Hamiltonian is
\be
\label{38}
\widetilde H = \bigoplus_f H_f(w_f) \equiv \widetilde H(w) \;  ,
\ee
in which
\be
\label{39}
 H_f(w_f) = \lim_{\xi_f\ra w_f} \; H_f(\xi_f) \;  .
\ee
The phase probabilities $w_f$ are the minimizers of the thermodynamic potential
\be
\label{40}
 \Om = {\rm abs}\; \min \Om(w) \;  ,
\ee
where
\be
\label{41}
 \Om(w) = - T \ln \; {\rm Tr} \left\{ - \bt \widetilde H(w)\right\} \; .
\ee
The minimization is accomplished under the normalization conditions
\be
\label{42}
 \sum_f w_f = 1 \; , \qquad 0 \leq \; w_f \; \leq 1 \;  .
\ee

\vskip 2mm

{\bf Theorem 3}. The observable quantities, defined by the averages 
\be
\label{43}
\lgl \; \hat A \; \rgl = 
{\rm Tr} \int \hat\rho(\xi) \hat A(\xi) \; \cD\xi 
\ee
of the operators
\be   
\label{44}
\hat A(\xi) = \bigoplus_f \hat A_f(\xi_f) \;  ,
\ee
with
\be
\label{45}
\hat A_f(\xi_f) = \sum_{m=0}^\infty \int \xi_f(\br_1) \xi_f(\br_2) \ldots
\xi_f(\br_m) \; A_f(\br_1,\br_2,\ldots,\br_m) \; d\br_1 d\br_2\ldots d\br_m
\ee
after the averaging over phase configurations, reduce to the form
\be
\label{46}
 \lgl \; \hat A \; \rgl = {\rm Tr}\; \hat\rho(w) \hat A(w)  \; ,
\ee
in which the renormalized operators are
\be
\label{47}
\hat A(w) = \bigoplus_f \hat A_f(w_f) \; ,
\ee
with
\be
\label{48}
 \hat A_f(w_f) = \sum_{m=0}^\infty \; w_f^m \int 
A_f(\br_1,\br_2,\ldots,\br_m) \; d\br_1 d\br_2\ldots d\br_m \; ,
\ee
and the renormalized statistical operator is
\be
\label{49}
\hat\rho(w) = \frac{1}{Z} \; \exp \left\{ - \bt \widetilde H(w) \right\} \; ,
\ee
with the partition function
\be
\label{50}
 Z = {\rm Tr} \; \exp \left\{ - \bt \widetilde H(w) \right\}\;  .
\ee

\vskip 2mm

{\bf Remark}. If the thermodynamic potential (\ref{29}), by expanding it in 
powers of the manifold indicator functions, can be represented in the form of 
the functional (\ref{31}), then the thermodynamic potential (\ref{30}) acquires 
the form of the thermodynamic potential (\ref{36}), however with the weights 
$w_f$ that are not the minimizers of the thermodynamic potential, but are given 
by the values prescribed by the corresponding volumes occupied  by the frozen 
phases.

\section{Geometric phase probabilities}

According to definition (\ref{34}), the weights $w_f$ are the geometric phase 
probabilities
\be
\label{51}
 w_f = \frac{V_f}{V} \;  .
\ee
Since
\be
\label{52}
\sum_f N_f = N \; , \qquad  \sum_f V_f = V \;  ,
\ee
probabilities (\ref{51}) meet the normalization conditions (\ref{42}). 

For illustration, let us consider the case of two coexisting phases, $f=1,2$. 
Then, to satisfy conditions (\ref{42}), we may set
\be
\label{53}
 w \equiv w_1 \; , \qquad w_2 = 1 - w \;  .
\ee

When $w_f$ are not identically $0$ or $1$, they are defined by minimizing the 
thermodynamic potential, so that
\be
\label{54}
 \frac{\prt\Om}{\prt w} = 0 \; , \qquad 
\frac{\prt^2\Om}{\prt w^2} > 0 \;  .
\ee
The first of these conditions gives
\be
\label{55}
 \left\lgl \; \frac{\prt\widetilde H}{\prt w} \; \right\rgl = 0 \;  ,
\ee
with the Hamiltonian (\ref{38}). 

Keeping in view Hamiltonians generated by form (\ref{7}), we have the replica 
Hamiltonians 
\be
\label{56}
 H_f(w_f) = \hat H_f(w_f) - \mu \hat N_f(w_f) \;  ,
\ee
with the energy Hamiltonians
$$
\hat H_f(w_f) = w_f \int \psi_f^\dgr(\br) \; \left[ \; 
- \; \frac{\nabla^2}{2m} + U(\br) \; \right] \; \psi_f(\br) \; d\br \; +
$$
\be
\label{57}
+ \;
\frac{w_f^2}{2} \int \psi_f^\dgr(\br) \psi_f^\dgr(\br')\; \Phi(\br-\br')\;
\psi_f(\br') \psi_f(\br) \; d\br d\br' 
\ee 
and the number-of-particle operators
\be
\label{58}
 \hat N_f(w_f) = w_f \int  \psi_f^\dgr(\br) \psi_f(\br) \; d\br \; .
\ee

Introducing the notations
$$
\hat K_f \equiv \int  \psi_f^\dgr(\br) \left[ \; 
- \; \frac{\nabla^2}{2m} + U(\br) \; \right] \; \psi_f(\br) \; d\br \; ,
$$
$$
\hat \Phi_f \equiv \int  \psi_f^\dgr(\br) \psi_f^\dgr(\br') \;
\Phi(\br-\br') \; \psi_f(\br') \psi_f(\br) \; d\br d\br' \; ,
$$
\be
\label{59}
 \hat R_f \equiv \int  \psi_f^\dgr(\br) \psi_f(\br) \; d\br \;  ,
\ee
makes it straightforward to represent Hamiltonians (\ref{56}) in the simple way
\be
\label{60}
 H_f(w_f) = w_f \hat K_f + \frac{w_f^2}{2} \; \hat \Phi_f -
\mu w_f \hat R_f \; .
\ee

As is seen, the replica Hamiltonians (\ref{60}) depend on $w_f$ explicitly. At 
the same time, they also depend on $w_f$ implicitly through the dependence of 
the field operators on $H_f(w_f)$, as far as the field operators satisfy the 
Heisenberg equations
\be
\label{61}
 i \; \frac{\prt}{\prt t} \; \psi_f(\br,t) = 
\left[\; \psi_f(\br,t), \; H_f(w_f)\; \right] \; .
\ee
The other equivalent form of these equations \cite{Yukalov_25,Yukalov_26} reads 
as
\be
\label{62}
 i \; \frac{\prt}{\prt t} \; \psi_f(\br,t) = 
\frac{\dlt H_f(w_f)}{\dlt\psi_f^\dgr(\br,t)} \;  .
\ee

In order to grasp the feeling of the structure of the equations for the phase 
probabilities, let us for a while neglect the implicit dependence on $w_f$, 
assuming that the explicit dependence prevails. Then Eq. (\ref{55}) yields
\be
\label{63}
 w = \frac{\Phi_2 + K_2 - K_1 +\mu(R_1-R_2)}{\Phi_1+\Phi_2} \;  ,
\ee
where
\be
\label{64}
 K_f = \lgl \; \hat K_f\; \rgl \; , \qquad  
\Phi_f = \lgl \; \hat \Phi_f\; \rgl \; , \qquad 
R_f = \lgl \; \hat R_f\; \rgl \;  .
\ee
The second of Eqs. (\ref{54}) yields the stability condition
\be
\label{65}
 \left\lgl \; \frac{\prt^2\widetilde H}{\prt w^2} \; \right\rgl = 
\bt \left\lgl \; 
\left( \frac{\prt\widetilde H}{\prt w} \right)^2 \; \right\rgl \; ,
\ee
for which the necessary condition is
\be
\label{66}
\Phi_1 + \Phi_2 ~ > ~ 0 \;  .
\ee

It is useful to connect the geometric phase probabilities with the typical 
densities in the system. The particle density of an $f$-th phase is given by 
the ratio
\be
\label{67}
  \rho_f \equiv \frac{N_f}{V_f} = \frac{R_f}{V} \;  .
\ee
The average density in the system is
\be
\label{68}
  \rho \equiv \frac{N}{V} = \sum_f w_f \rho_f  \;   .
\ee
Th fraction of particles in an $f$-th phase reads as
\be
\label{69}
 n_f \equiv \frac{N_f}{N} \; ,
\ee
with the evident conditions
\be
\label{70}
\sum_f n_f  = 1 \; , \qquad 0 \leq ~ n_f ~ \leq 1 \;  .
\ee
From the relation
\be
\label{71}
 w_f \rho_f = n_f \rho \; ,
\ee
it follows
\be
\label{72}
 \sum_f \rho_f = \rho \sum_f \frac{n_f}{w_f} \; .
\ee

The situation simplifies, when the phases are distinguished not by their 
densities, but by some other properties, like magnetic, electric, or 
orientational features, while the phase densities are equal. In that case, 
the phase probabilities coincide with the phase fractions,
\be
\label{73}
 w_f = n_f \qquad (\rho_f = \rho ) \;  .
\ee
And the relations
\be
\label{74}
R_f = \frac{N_f}{w_f} = \frac{N_f}{n_f} = N
\ee
hold. Then in the equation for the phase probability (\ref{63}), the term with 
the chemical potential vanishes.

\section{Classical heterophase systems}

For generality, we have considered above quantum systems. Of course, heterophase 
systems do not need to be necessarily quantum, but classical systems can also 
be heterophase. In the present section, we show how the theory is applied to 
classical systems.

Classical $N$ particle systems are characterized by the position coordinates 
and momentum variables
\be
\label{75}
 q = \{ \bq_1,\bq_2,\ldots,\bq_N\} \; , \qquad
 p = \{ \bp_1,\bp_2,\ldots,\bp_N\} \; .
\ee
For concreteness, three-dimensional spaces of position and momentum coordinates 
are kept in mind, so that there are in total $6N$ variables in the {\it space of 
microstates} $\{q,p\}$. In that space, a measure $\mu(q,p)$ is given, making it 
a {\it measurable phase space}, or simply a {\it phase space}
\be
\label{76}
 \mathbb{G} = \left\{q,p,\mu(q,p) \right\} \;  .
\ee

Classical mechanics usually is equipped with the differential measure
\be
\label{77}
 d\mu(q,p) = \frac{dq\; dp}{N!(2\pi\hbar)^{3N} } \;  .
\ee
Recall that the Planck constant $\hbar$ is included here in order to make the measure 
dimensionless, which is necessary for making dimensionless the effective number of 
states 
$$
W(E) = \int_{H(q,p)<E} d\mu(q,p)
$$
in the phase volume bounded by the energy surface corresponding to the energy $E$. 
This, in turn, is required for making dimensionless the expression under the logarithm 
in the Boltzmann formula for entropy 
$$
S = k_B \; \ln \; W(E) \;  .
$$
The use of $\hbar$ for the purpose of making dimensionless the phase-space measure 
and the effective number of states is dictated by the necessity of guaranteeing 
smooth transition between quantum and classical statistics. This inclusion of $\hbar$ 
in classical statistical mechanics is the standard commonly employed method that does 
not influence the expressions of either observable quantities or thermodynamic 
characteristics (see textbooks, e.g. \cite{Huang_51,Kubo_52,Isihara_53}), but correctly 
defines the Boltzmann entropy.   

Suppose that the system volume $V$ is filled by a mesoscopic mixture of several 
thermodynamic phases enumerated by $f=1,2,\dots$ and that are distinguished by 
different order parameters or order indices \cite{Yukalov_27,Yukalov_28}. The 
spatial location of these phases is characterized by the manifold indicator 
functions $\xi_f(\br)$, defined in Eq. (\ref{2}), for which
\be
\label{78}
 \int \xi_f(\br)\; d\br = V_f \; , \qquad \sum_f V_f = V \;  .
\ee
For each phase, there exists a probability density $\rho_f(q,p,\xi_f)$ normalized 
as
\be
\label{79}
 \int \rho_f(q,p,\xi_f)\; d\mu(q,p) = 1 \;  .
\ee
The phase space, complemented by the probability density, composes a 
{\it statistical ensemble}
\be
\label{80}
 \mathbb{E}_f = \left\{ \mathbb{G},\rho_f(q,p,\xi_f) \right\} \;  .
\ee
On a statistical ensemble (\ref{80}), the representatives of observables 
$A_f(q,p,\xi_f)$ are defined, whose averages yield the observable quantities
\be
\label{81}
 \lgl \; A_f \; \rgl = 
\int A_f(q,p,\xi_f) \rho_f(q,p,\xi_f) \; \cD\xi \; d\mu(q,p) \;  .
\ee

The statistical ensemble of the whole system is the Cartesian product
\be
\label{82}
 \mathbb{E} = {\rm x}_f \mathbb{E}_f = 
\left\{ \mathbb{G},\rho(q,p,\xi) \right\} \;  ,
\ee
with the density distribution
\be
\label{83}
 \rho(q,p,\xi) = {\rm x}_f  \rho_f(q,p,\xi_f) \;   .
\ee
The representatives of observables for the system are
\be
\label{84}
 A(q,p,\xi) = \bigoplus_f   A_f(q,p,\xi) \; ,
\ee
so that the system observable quantities read as
\be
\label{85}
\lgl \; A \; \rgl = \int A(q,p,\xi) \rho(q,p,\xi) \cD\xi \; d\mu(q,p) \; .
\ee

For example, the system Hamiltonian, similarly to Eq. (\ref{5}), is 
\be
\label{86}
H(q,p,\xi) = \bigoplus_f H_f(q,p,\xi_f) \;   ,
\ee
in which the phase-replica Hamiltonians are
\be
\label{87}
 H_f(q,p,\xi_f) = \sum_{i=1}^N \xi_f(\br_i) \left[\; \frac{\bp_i^2}{2m} +
U(\br_i) \; \right] + \frac{1}{2} 
\sum_{i\neq j}^N \xi_f(\br_i) \xi_f(\br_j) \Phi(\br_i-\br_j) \;  .
\ee

The density distribution of the system is normalized,
\be
\label{88}
\int \rho(q,p,\xi) \cD\xi \; d\mu(q,p) = 1 \;  .
\ee
The average system energy is
\be
\label{89}
 \int \rho(q,p,\xi) H(q,p,\xi)\; \cD\xi \; d\mu(q,p) = E \;  .
\ee
The number of particles in the system, $N$, is fixed. 

The information functional acquires the form
$$
I[\;\rho\;] = \int \rho(q,p,\xi)\; 
\ln\; \frac{\rho(q,p,\xi)}{\rho_0(q,p,\xi)}\; \cD\xi \; d\mu(q,p) \; + 
$$
\be
\label{90}
 +\;
\al \left[ \;  \int \rho(q,p,\xi)\; \cD\xi \; d\mu(q,p)\; - \; 
1 \;\right] \; + \; 
\bt \left[ \;  \int \rho(q,p,\xi)\; H(q,p,\xi)\; \cD\xi \; d\mu(q,p)\; - \; 
E \;\right] \; .
\ee
The minimization of this functional, in the case of no preliminary information, 
gives
\be   
\label{91}
\rho(q,p,\xi) = \frac{1}{Z} \; 
\exp \left\{ -\bt H(q,p,\xi) \right\} \;   ,
\ee
with the partition function 
\be
\label{92}
 Z = \int  
\exp \left\{ -\bt H(q,p,\xi) \right\} \; \cD\xi \; d\mu(q,p) \; .
\ee

After averaging over phase configurations, as is described in the previous 
sections, we come to the renormalized Hamiltonian
\be
\label{93}
H(q,p,w) = \bigoplus_f H_f(q,p,w_f) \;  ,
\ee
with the phase-replica terms
\be
\label{94}
H_f(q,p,w_f) = w_f 
\sum_{i=1}^N \left[\; \frac{\bp_i^2}{2m} + U(\br_i) \; \right] \; + \;
\frac{w_f^2}{2} \sum_{i\neq j}^N \Phi(\br_i-\br_j) \; .
\ee
Respectively, the functions representing observable quantities become 
\be 
\label{95}
A(q,p,w) = \bigoplus_f A_f(q,p,w_f) \;  .
\ee

The distribution function reads as
\be
\label{96}
\rho(q,p,w) = {\rm x}_f \rho_f(q,p,w_f) \;  ,
\ee
whose factors are
\be
\label{97}
\rho_f(q,p,w_f) = \frac{1}{Z_f} \; 
\exp \left\{ -\bt H_f(q,p,w_f) \right\} \;   ,
\ee
with the normalization
\be
\label{98}
\int \rho_f(q,p,w_f) \; d\mu(q,p) = 1
\ee
and the partition functions
\be
\label{99}
 Z_f = \int \exp \left\{ -\bt H_f(q,p,w_f) \right\} \; d\mu(q,p) \; .
\ee
Integrating over momenta gives
\be
\label{100}
 Z_f = \left( \frac{m k_B T}{2\pi\hbar w_f}\right)^{3N/2} \;
\int \exp\left\{ - \; \frac{w_f^2}{2k_B T} \sum_{i\neq j}^N \Phi(\br_i-\br_j)\;
- \; \frac{w_f}{k_B T} \sum_{i=1}^N U(\br_i) \right\} \; \frac{dq}{N!} \; .
\ee

The system free energy takes the form
\be
\label{101}
 F = \sum_f F_f = F(w) \;  ,
\ee
with the terms
\be
\label{102}
 F_f = - T \ln \; Z_f \; .
\ee
The phase probabilities are the minimizers of the free energy,
\be
\label{103}
F = {\rm abs}\;\min F(w) \; .
\ee

The observable quantities are the averages
\be
\label{104}
 \lgl \; A \; \rgl = \sum_f \; \lgl \; A_f \; \rgl \; ,
\ee
where
\be
\label{105}
\lgl \; A_f \; \rgl = 
\int \rho_f(q,p,w_f) A_f(q,p,w_f) \; d\mu(q,p) \; .
\ee

\section{Quasiaverages in classical statistics}

In order to distinguish different thermodynamic phases of quantum systems, 
there are several methods, such as the Frenkel \cite{Frenkel_29}, method of 
restricted phase space that is a classical analog of the Brout \cite{Brout_30} 
method of restricted trace for quantum systems. The idea of these methods is to 
integrate not over all phase space, that is, over the whole range of spatial 
coordinates and momentum variables, but to limit the integration over a 
restricted region of the phase space, such that would provide the description 
for the required thermodynamic phase. Details of this approach can be found 
in the review article \cite{Yukalov_7}. 

In a more general picture, it is possible to define {\it weighted phase 
spaces}, following the idea of introducing {\it weighted Hilbert spaces} 
\cite{Yukalov_7,Yukalov_10}. For this purpose, the differential measure of 
a phase space is weighted with an auxiliary distribution $\varphi_f(q,p)$, 
so that this measure becomes
\be
\label{106}
 d\mu_f(q,p) = \frac{dq dp}{N!(2\pi\hbar)^{3N}}\; \vp_f(q,p) \; .
\ee
The auxiliary distribution weights the points of the phase space so that to obtain 
the description characterizing the $f$-th thermodynamic phase. The weighted phase 
space for an $f$-th phase is
\be
\label{107}
 \mathbb{G}_f = \{q,p,\mu_f(q,p)\} \; .
\ee
The phase space of a heterophase classical system takes the form
\be
\label{108}
  \mathbb{G} = {\rm x}_f  \mathbb{G}_f \;  .
\ee
In the case where the auxiliary distribution is either absent or a constant, 
the method of weighted phase space reduces to the method of restricted phase 
space.

Technically, the selection of a phase space, required for a correct description 
of a needed thermodynamic phase, can be done by imposing constraints, such as 
symmetry breaking, on the corresponding distribution (\ref{97}). A convenient 
way of symmetry breaking is by introducing infinitesimal sources, as was 
mentioned by Kirkwood \cite{Kirkwood_31} and developed by Bogolubov 
\cite{Bogolubov_32,Bogolubov_33,Bogolubov_34} into the method of quasiaverages.   

Let us illustrate this for the case of distinguishing a periodic crystalline 
phase from a uniform disordered phase, being based on the Hamiltonians (\ref{93}) 
and (\ref{94}). For concreteness, let us label the periodic crystalline phase 
by $f=1$, while the uniform phase by $f=2$. As an order characteristic, one can 
accept the particle density equipped with the related symmetry properties. For 
the mixture of two phases, the representative of the observable density is
\be
\label{109}
\hat\rho(\br,\xi) =  
\hat\rho_1(\br,\xi_1) \; \bigoplus \; \hat\rho_2(\br,\xi_2) \; ,
\ee
where
\be
\label{110}
\hat\rho_f(\br,\xi_f) =  \sum_{i=1}^N \xi_f(\br) \dlt(\br-\br_i) \; .
\ee
After averaging over phase configurations, we have
\be
\label{111}
\hat\rho(\br,w) = 
\hat\rho_1(\br,w_1) \; \bigoplus \; \hat\rho_2(\br,w_2) \; , 
\ee  
with
\be
\label{112}
 \hat\rho_f(\br,w_f) = w_f \sum_{i=1}^N \dlt(\br-\br_j) \; .
\ee
The observable particle density reads as
\be
\label{113}
\rho(\br,w) = \lgl\; \hat\rho(\br,w)\;\rgl = 
\rho_1(\br,w_1) + \rho_2(\br,w_2) \;  ,
\ee
where
\be
\label{114}
\rho_f(\br,w_f) = 
\int \rho_f(q,p,w_f) \hat\rho_f(\br,w_f) \; d\mu(q,p) \; .
\ee
Integrating out the momentum variables results in the expression
\be
\label{115}
\rho_f(\br,w_f) = 
w_f \int \sum_{i=1}^N \dlt(\br-\br_i) g_f(q,w_f) \; \frac{dq}{N!} \; ,
\ee
in which
\be
\label{116}
g_f(q,w_f) = \frac{1}{Q_f} \exp\left\{ -\; \frac{w_f^2}{2k_B T}
\sum_{i\neq j}^N \Phi(\br_i-\br_j) \; -\; 
\frac{w_f}{k_B T} \sum_{i=1}^N U(\br_i) \right\} \; \; .
\ee
  
In the absence of any external potential, when $U(\br)\equiv 0$, the integral 
in the right-hand side of Eq. (115) gives a constant value, which is okay for 
the uniform phase but is not suitable for the crystalline phase. To overcome 
this problem in the case of a crystalline phase, it is possible to set
\be
\label{117}
 U(\br) = \ep U_L(\br) \; ,
\ee
assuming a lattice potential periodic over the appropriate crystalline lattice,
\be
\label{118}
  U_L(\br + \ba) =  U_L(\br) \;  .
\ee
Here $\ep$ is a small parameter. To stress that this parameter enters the 
expression of the related density, we shall denote the latter as 
$\rho_1(\br,w_1,\ep)$. Also, let us recall the definition of the thermodynamic 
limit
\be
\label{119}
N ~ \ra ~ \infty \; , \qquad V ~ \ra ~ \infty \; , \qquad 
\frac{N}{V} ~ \ra ~ const \; .
\ee
The correct density, periodic over the crystalline lattice, is given by the 
limiting procedure producing the quantity
\be
\label{120}
 \rho_1(\br,w_1) = \lim_{\ep\ra 0} \;\lim_{N\ra\infty} \; \rho_1(\br,w_1,\ep)
= \rho_1(\br + \ba,w_1)  
\ee
called a quasiaverage \cite{Bogolubov_32,Bogolubov_33,Bogolubov_34}. At the 
same time, by setting $U_L(\br)\equiv 0$, one obtains a uniform density
\be
\label{121}
\rho_2(\br,w_2) = \rho_2 = const \qquad ( U(\br) \equiv 0 )
\ee
corresponding to a disordered phase.  
 
Instead of the double limiting procedure (\ref{120}), it is possible to define 
a single limiting procedure by employing the source
\be
\label{122}
U(\br) = \frac{1}{N^\gm}\; U_L(\br) \qquad ( 0 < \gm < 1 )
\ee
called thermodynamic quasiaverage \cite{Yukalov_35,Yukalov_36}.

In any case, the introduction of an auxiliary potential, is equivalent to the 
definition of a weighted phase space with the auxiliary distribution
\be
\label{123}
 \vp_1(q,\ep) = \exp\left\{ - \; 
\frac{w_f}{k_B T} \sum_{i=1}^N \ep U_L(\br_i) \right\} \; .
\ee

\section{Surface free energy}

When on microscopic level, there is a spatial phase separation, then on 
macroscopic level, there appears the concept of the surface free energy and 
the related thermodynamic quantities. This also happens in the case of the 
mesoscopic phase separation.

The surface free energy is defined 
\cite{Ono_37,Rusanov_38,Rusanov_39,Kjelstrup_40,Bedeaux_41} as the difference 
between the actual free energy of the system and the sum of the free energies 
of macroscopically separated pure Gibbs phases,
\be
\label{124}
 F_{sur} = F(w) - F_G \; .
\ee
Here the free energy of a heterophase system with mesoscopic nanosize phase 
separation in two intermixed phases, according to the above theory, has the 
form
\be
\label{125}
 F(w) = F_1(w_1) + F_2(w_2) \;  .
\ee
But the separation into two pure Gibbs phases, occupying the volumes $V_1$ 
and $V_2$, leads to the free energy
\be
\label{126}
 F_G = F_1^G + F_2^G = w_1 F_1(1) + w_2 F_2(1) \;  .
\ee
In this way, the surface free energy for a heterophase mixture of two phases is
\be
\label{127}
 F_{sur} = F_1(w_1) +  F_2(w_2) - w_1 F_1(1) - w_2 F_2(1) \; .
\ee

The Gibbs macroscopic mixture of pure phases (\ref{126}) corresponds to a linear 
combination of the related free energies of pure phases. However, the effective 
free energy of a mesoscopic mixture (\ref{125}) is not a linear combination of 
the pure-phase free energies. As has been shown in the above sections, the free 
energy of a heterophase mixture is only by the form looks as a sum of two terms. 
However these terms correspond not to pure phases, but to effective renormalized 
expressions nonlinearly depending on phase probabilities.

\section{Crystal with regions of disorder}

To illustrate the above theory, let us consider the model of a crystalline 
solid with nanosize regions of disorder. As examples, we can keep in mind 
solids with pores and cracks \cite{Yukalov_42,Yukalov_43,Yukalov_44}, crystals 
with dislocations \cite{Langer_17,Langer_18,Langer_19}, optical lattices with 
regions of broken periodicity \cite{Yukalov_45}, crystals with amorphous 
inclusions \cite{Bakai_46}, and quantum crystals with vacancy clusters 
\cite{Boninsegni_47,Ma_48,Singh_49,Yukalov_50}.

The ordered and disordered phases are distinguished by their densities
\be
\label{128}
 \rho_f \equiv \frac{N_f}{V_f} = \frac{1}{V} \int \lgl \; \psi_f^\dgr(\br)
\psi_f(\br) \; \rgl \; d\br \;  ,
\ee
so that the density of the ordered phase $\rho_1$ is larger than the density 
$\rho_2$ 
of the disordered phase:
\be
\label{129}
 \rho_1 ~ > ~ \rho_2 \;  .
\ee  

Following the theory expounded above, the renormalized grand Hamiltonian of 
a two phase mixture has the form
\be
\label{130}
 \widetilde H = H_1 \; \bigoplus \; H_2 \;  ,
\ee
with the replica Hamiltonians
$$
H_f = w_f \int \psi_f^\dgr(\br) \left[ \; 
\hat H_L(\br) - \mu \right] \psi_f(\br) \; d\br \; +
$$
\be
\label{131}
+ \;
\frac{w_f^2}{2} \int \psi_f^\dgr(\br) \psi_f^\dgr(\br')\; \Phi(\br-\br') \;
\psi_f(\br') \psi_f(\br)\; d\br d\br' \; ,
\ee
where 
$$
\hat H_L(\br) = -\; \frac{\nabla^2}{2m} + U(\br) \;  .
$$
We keep in mind a solid state with well localized particles, both in the 
ordered crystalline phase as well as in the disordered phase. Therefore the 
field operators can be expanded over localized orbitals $\vp_{nj}$ as
\be
\label{132}
\psi_f(\br) = \sum_{nj} \; e_{jf} c_{nj} \vp_{nj}(\br) \;  ,
\ee
where the index $j=1,2,\ldots,N_L$ enumerates lattice sites, $n$ is the index 
of quantum states, $c_{nj}$ is an annihilation operator of a particle in the 
state $n$ at the lattice site $j$, and $e_{jf}$ equals one or zero depending 
on whether the site $j$ is occupied or free.       

We assume that particles are well localized in their lattice sites, so that 
their hopping between different sites can be neglected, and that each lattice 
site can be occupied not more than by one particle. These conditions imply 
the unipolarity properties
\be
\label{133}
 \sum_n c_{nj}^\dgr c_{nj} = 1 \; , \qquad c_{mj} c_{nj} = 0 \; .
\ee
The absence of hopping between the lattice sites means that only diagonal 
matrix elements survive,
\be
\label{134}
E_0 \equiv \lgl \; nj \; | \; \hat H_L \; | \; jn \;\rgl \; , \qquad
\Phi_{ij} \equiv \lgl\; ni, nj \; | \; \Phi \; | \; jn, in \; \rgl \;  .
\ee
The constant term $E_0$ can be included in the chemical potential. As a result, 
substituting the field-operator expansion (\ref{132}) into Hamiltonian (\ref{131}) 
yields
\be
\label{135}
 H_f = 
\frac{w_f^2}{2} \sum_{i\neq j}^{N_L} \Phi_{ij} e_{if} e_{jf} \; - \;
\mu w_f \sum_{j=1}^{N_L} e_{jf} \; .
\ee

Resorting to the canonical transformation 
$$
e_{jf} = \frac{1}{2} + S_{jf}^z \qquad ( e_{jf} = 0,1 ) \; ,
$$
\be
\label{136}
S_{jf}^z = e_{jf} \; - \; \frac{1}{2}  \qquad 
\left( S_{jf}^z = \pm \frac{1}{2} \right) \; ,
\ee
the above Hamiltonian can be rewritten in the pseudospin representation
$$
H_f = \frac{N_L}{8} \; \left( w_f^2 \Phi - 4 w_f \mu \right) \; + 
$$
\be
\label{137}
+ \;
\frac{1}{2}\; \left( 
w_f^2 \Phi - 2 w_f \mu \right) \sum_{j=1}^{N_L} S_{jf}^z \; + \;
\frac{w_f^2}{2} \sum_{i\neq j}^{N_L} \Phi_{ij} S_{if}^z S_{jf}^z \; ,
\ee    
in which the parameter 
\be
\label{138}
 \Phi \equiv \frac{1}{N_L} \sum_{i\neq j}^{N_L} \Phi_{ij} ~ > ~ 0  
\ee
has to be positive in view of the stability condition (\ref{66}).

We use the mean-field approximation
\be
\label{139}
S_{if}^z S_{jf}^z = S_{if}^z \lgl\; S_{jf}^z \; \rgl \; + \;
\lgl\; S_{if}^z \; \rgl S_{jf}^z \; - \; 
\lgl\; S_{if}^z \; \rgl \lgl\; S_{jf}^z \; \rgl
\ee
and introduce the notation
\be
\label{140}
s_f \equiv 2 \lgl\; S_{jf}^z \; \rgl = 
\frac{2}{N_L} \sum_{j=1}^{N_L} \lgl\; S_{jf}^z \; \rgl \; .
\ee
Then we have the relations
\be
\label{141}
 s_f = 2 \lgl\; e_{jf} \; \rgl \; - \; 1 \; , \qquad
 \lgl\; e_{jf} \; \rgl = \frac{1+s_f}{2} \; .
\ee

The density of the $f$-th phase (\ref{128}) has the form
\be
\label{142}
 \rho_f = \frac{1}{V} \sum_{j=1}^{N_L} \lgl\; e_{jf} \; \rgl \; .
\ee
It is convenient to define the dimensionless density fractions
\be
\label{143}
  x_f \equiv \frac{1}{N_L} \sum_{j=1}^{N_L} \lgl\; e_{jf} \; \rgl \; .
\ee
Introducing the lattice filling factor
\be
\label{144}
\nu \equiv \frac{N}{N_L}
\ee
transforms the density fractions (\ref{143}) into
\be
\label{145}
 x_f = \frac{\rho_f}{\rho} \; \nu = \frac{1+s_f}{2} \;  .
\ee

For the grand thermodynamic potential we obtain
$$
\frac{\Om}{N_L} = 
\frac{1}{8} \sum_f w_f^2 ( 1 - s_f^2) \Phi \; - \; 2T\ln 2 \; - \;
\frac{\mu}{2} \; - 
$$
\be
\label{146}
-\;
T \sum_f \ln \; \cosh \left[ \; 
\frac{w_f^2(1+s_f)\Phi-2w_f\mu}{4T} \; \right] \; .
\ee
Minimizing the thermodynamic potential with respect to $s_f$ gives
\be
\label{147}
 s_f = \tanh 
\left[ \; \frac{2w_f\mu - w_f^2(1+s_f)\Phi}{4T}\; \right] \;  .
\ee
And minimizing with respect to $w_f$, under the normalization condition
\be
\label{148}
 w_1 + w_2 = 1 \; ,
\ee
results in the equation
\be
\label{149}
 \frac{\mu}{\Phi} = \frac{w_1(1+s_1)^2-w_2(1+s_2^2)}{2(s_1-s_2)} \;  .
\ee
Using the relation
\be
\label{150}
 \frac{N_f}{N_L} = w_f x_f = \frac{w_f}{2}\; ( 1 + s_f ) \; ,
\ee
for the filling factor we have
\be
\label{151}
 \nu = \frac{N}{N_L} = \frac{N_1+N_2}{N_L} = 
\frac{1}{2} \sum_f w_f (1 + s_f) \;  .
\ee
From the latter it follows
\be
\label{152}
 w_1 = \frac{2\nu-1-s_2}{s_1-s_2} \; , \qquad  
w_2 = \frac{2\nu-1-s_1}{s_2-s_1} \;   .
\ee

The role of the order parameters is played by the densities (\ref{129}) or 
by the dimensionless density fractions (\ref{143}). According to condition 
(\ref{129}), distinguishing the ordered and disordered phases, 
\be
\label{153}
 x_1 ~ > ~ x_2 \;  .
\ee
Taking into account that the filling factor (\ref{151}) can be written in 
the form
\be
\label{154}
\nu = w_1 x_1 + w_2 x_2
\ee
and that $0 \leq w_f \leq 1$, it follows that
\be
\label{155}
 0 ~ \leq ~ x_2 ~ \leq \nu ~ \leq ~ x_1 ~ \leq 1 \;  .
\ee
 
To analyze the system stability, it is necessary to minimize the thermodynamic 
potential. For convenience, we define the dimensionless free energy
\be
\label{156}
 F \equiv \frac{\Om+\mu N}{N_L \Phi}  
\ee
and the related dimensionless specific heat and isothermal compressibility
\be
\label{157}
 C_V = - T\left( \frac{\prt^2F}{\prt T^2}\right)_V \; , \qquad
\kappa_T = \frac{1}{\nu^2} \left( \frac{\prt^2F}{\prt\nu^2}\right)^{-1}_T \; .
\ee
Stability conditions require that the latter quantities satisfy the inequalities
\be
\label{158}
0 ~ \leq ~ C_V ~ < ~ \infty \; , \qquad 
0 ~ \leq ~ \kappa_T ~ < ~ \infty \; .
\ee

\section{System existence and stability}

To prove that the described above solid system presenting a crystal with 
nanoscopic regions of disorder really can exist and can be stable, we need 
to accomplish numerical investigations. For the following, it is useful to 
simplify the notation by setting for the ordered phase
\be
\label{159}
x_1 \equiv x \; , \qquad w_1 \equiv w
\ee
and for the disordered phase 
\be
\label{160}
x_2 \equiv y \; , \qquad w_2 \equiv 1 - w \;  .
\ee
Then the inequalities (\ref{155}) become
\be
\label{161}
 0 ~ \leq ~ y ~ \leq \nu ~ \leq ~ x ~ \leq 1 \;  .
\ee
The phase probabilities reduce to 
\be
\label{162}
 w = \frac{\nu-y}{x-y} \; , \qquad  1 - w = \frac{x-\nu}{x-y} \;  .
\ee
For the chemical potential (\ref{149}), we get
\be
\label{163}
  \frac{\mu}{\Phi} = \frac{w(x^2+y^2)-y^2}{x-y} \; .
\ee

Also, let us measure temperature in units of $\Phi$. Then the free energy 
(\ref{156}) takes the form
$$
F = \frac{1}{2} \left[\; 
w^2 x ( 1 - x) + ( 1 - w)^2 y ( 1 - y) \; \right] \; +
$$
\be
\label{164}
+ \;
\frac{T}{2} \; \ln \; \left[ x ( 1 - x) y ( 1 - y) \; \right] \; + \; 
\left( \nu \; - \; \frac{1}{2}\right) \; \frac{wx^2 - (1-w)^2y^2}{x-y} \; .
\ee
And for the order parameters (\ref{147}), we obtain
$$
2x = 1 + \tanh\left\{ \frac{wy\;[\;wx-(1-w)y\;]}{2T(x-y)}\right\} \; , 
$$
\be
\label{165} 
2y = 1 + \tanh\left\{ \frac{(1-w)x\;[\;wx-(1-w)y\;]}{2T(x-y)}\right\} \; .
\ee

The system of equations (\ref{162}) and (\ref{165}) is solved numerically under 
restriction (\ref{161}) and stability conditions (\ref{158}). The solution exists 
for the filling factor in the interval
\be
\label{166}
  0 ~ < ~ \nu ~ < ~ \frac{1}{2} 
\ee
and in the temperature range
\be
\label{167}
 T_n ~ < ~ T ~ < ~ T_n^* \;  .
\ee
The temperature $T_n \geq 0$ shows where the heterophase system appears, because 
of which this is called the lower nucleation temperature. When temperature rises, 
the heterophase state disappears at the upper nucleation temperature $T^*_n$ 
where the probability $w(T_n^*)$ becomes zero, which gives
\be
\label{168}
T_n^* = \frac{\nu}{(1-2\nu)[\;\ln(1-\nu) - \ln\nu\;]} \;  .
\ee
Table 1 shows the nucleation temperatures in-between which the crystal state 
with regions of disorder can exist. 

In the region of existence of the heterophase state, the free energy (\ref{164}) 
is lower than the free energy of the pure ordered state, where $w \equiv 1$. 
Therefore such a system with the regions of disorder, describing extended defects, 
in the intervals of the filling factor (\ref{166}) and temperature (\ref{167}), 
corresponds to a stable heterophase system.

\section{Conclusion}

The description of structures with randomly distributed extended defects is 
notoriously difficult, since such materials are heterophase and strongly 
nonuniform. However, the description can be simplified when one is interested 
in the averaged properties of a sample as a whole. In that case, it is possible 
to develop an approach based on the averaging over phase configurations. As a 
result, it is possible to reduce the consideration to the study of the replicas 
of the separate phases, described by effective renormalized Hamiltonians. As a 
price for this simplification, one comes to the necessity of dealing with these 
more complicated effective Hamiltonians and to the need of calculating the
geometric weights, or geometric probabilities of the phases. Nevertheless, it 
is worth paying the price because the overall problem becomes treatable, while 
the calculational complications are rather minor. The approach is illustrated 
by a lattice model with regions of disorder. It is shown that such nonideal 
structures can exist as stable statistical systems.    

It is important to emphasize that the developed approach can describe metastable 
as well as unstable states of matter. The system of equations, defining the order 
parameters and phase probabilities, is usually strongly nonlinear and displays 
several solutions. The solutions, for which the stability conditions, such as the 
positivity of specific heat, compressibility, and of other available susceptibilities, 
are not valid, correspond to unstable states. The solutions for which these stability 
conditions are satisfied, but the related free energy does not define an absolute 
minimum, describe metastable states. The solutions, leading to the lowest free energy 
and satisfying all stability conditions, correspond to stable states. Such a situation 
happens for the example treated in Secs. 10 and 11. This is why, it has been stressed 
that we need to choose those solutions that satisfy the stability conditions (\ref{158}) 
and minimize the free energy (\ref{156}). Only these solutions describe stable states, 
while other solutions, for which some of the stability conditions are broken, or free 
energy is not minimal, correspond to unstable or metastable states.

\vskip 2cm

\begin{table}[hp]
\caption{Lower and upper nucleation temperatures (in units of $\Phi$) for
different filling factors.}
\vskip 5mm
\centering
\renewcommand{\arraystretch}{1.2}
\begin{tabular}{|l|l|l|} \hline
$\nu$        &  $T_n$  &  $T_n^*$  \\ \hline
0.1          & 0       & 0.0157    \\ 
0.2          & 0       & 0.240     \\ 
0.3          & 0       & 0.885     \\ 
0.329        & 0.01    & 1.344     \\
0.330        & 0.0125  & 1.371     \\ 
0.40         & 0.521   & 4.933     \\
0.45         & 0.958   & 22.425    \\ \hline
\end{tabular}
\end{table}

\newpage

\begin{figure}[ht]
\centerline{
\hbox{ \includegraphics[width=8cm]{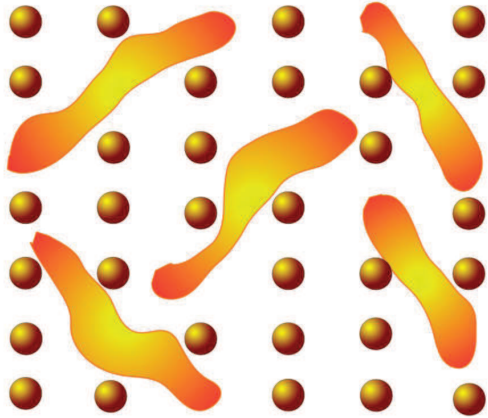}  } }
\caption{
Crystal with randomly distributed extended nanosize defects.
}
\label{fig:Fig.1}
\end{figure}

\vskip 3cm 

\begin{figure}[ht]
\centerline{
\hbox{ \includegraphics[width=8cm]{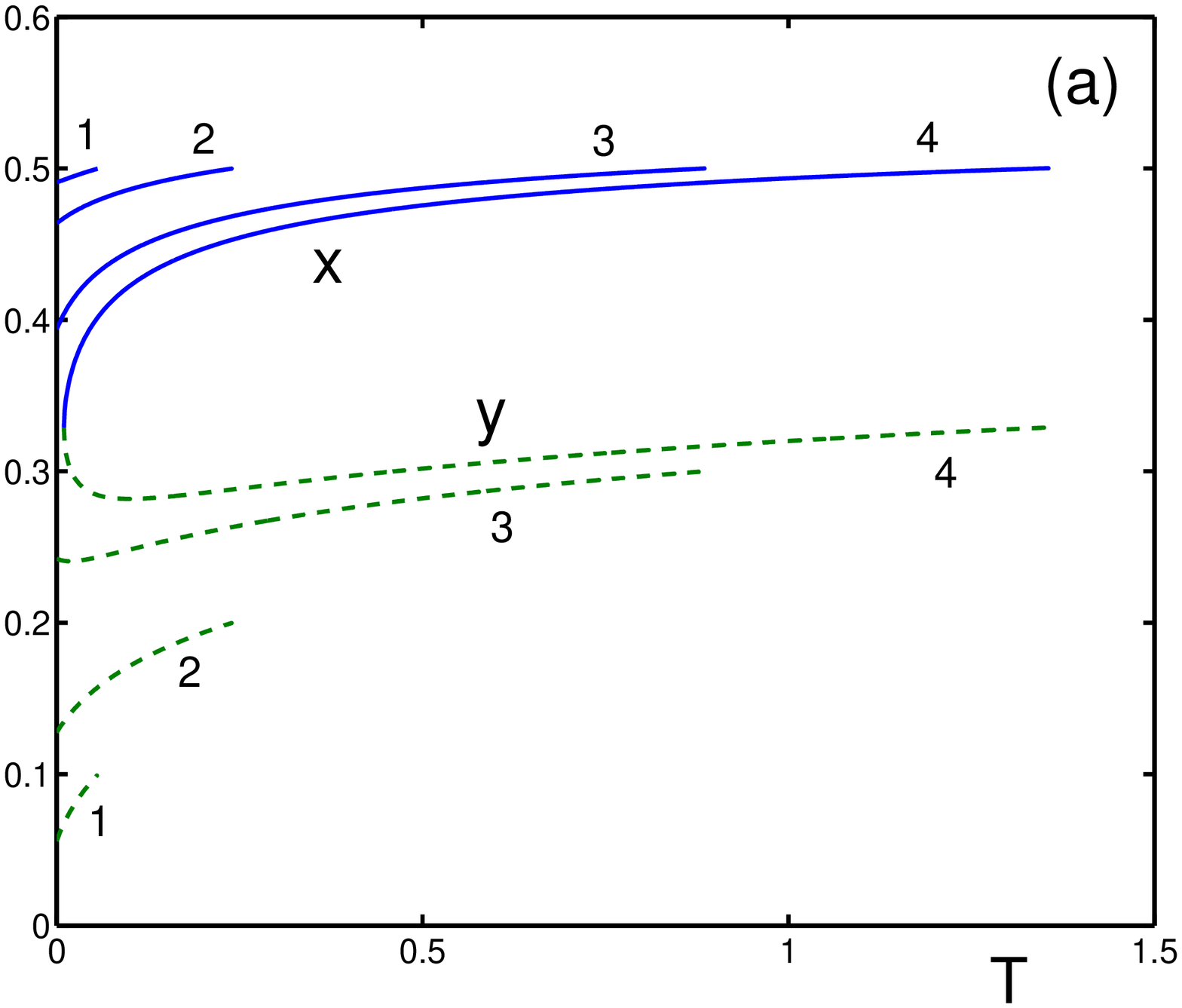} \hspace{1cm}
\includegraphics[width=8cm]{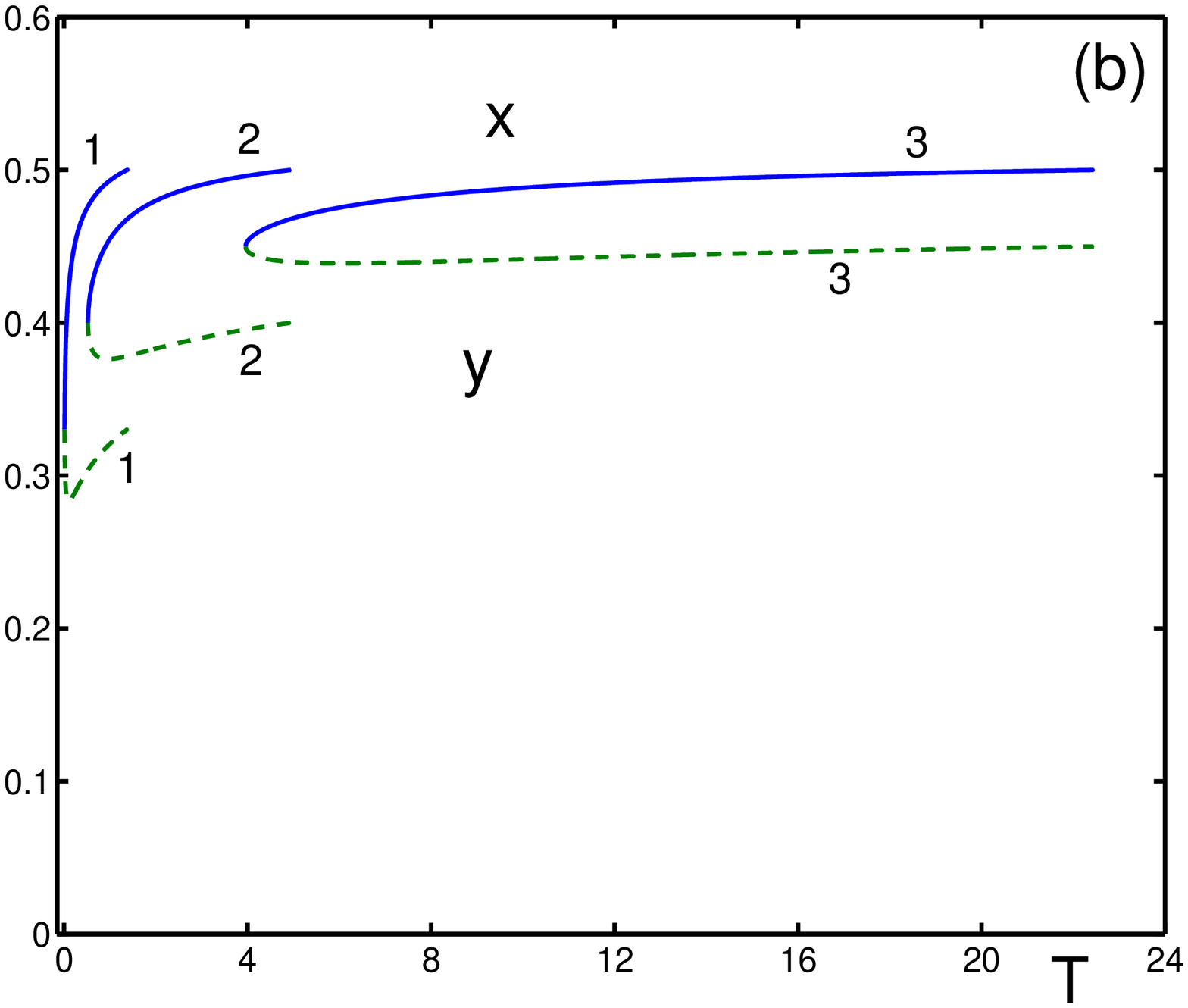}  } }
\caption{
Dimensionless densities of the ordered phase $x(t)$ (solid line) and of 
the disordered phase $y(t)$ (dashed line) as functions of dimensionless
temperature for different lattice filling factors:
(a) $\nu=0.1$ (line 1),  $\nu=0.2$ (line 2), $\nu=0.3$ (line 3), and
$\nu=0.329$ (line 4);
(b) $\nu=0.33$ (line 1),  $\nu=0.4$ (line 2), and $\nu=0.45$ (line 3).
}
\label{fig:Fig.2}
\end{figure}

\vskip 3cm

\begin{figure}[ht]
\centerline{
\hbox{ \includegraphics[width=8cm]{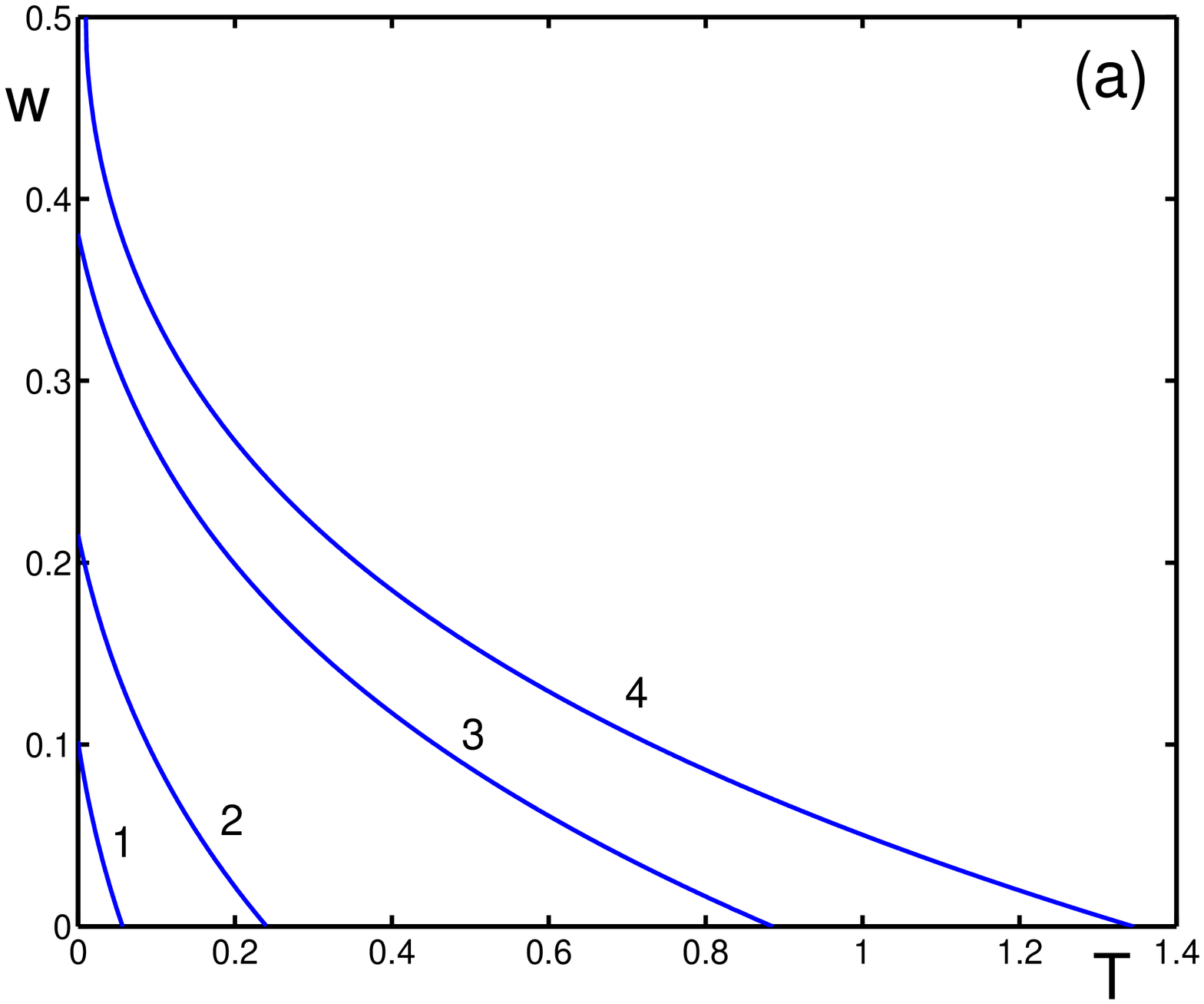} \hspace{1cm}
\includegraphics[width=8cm]{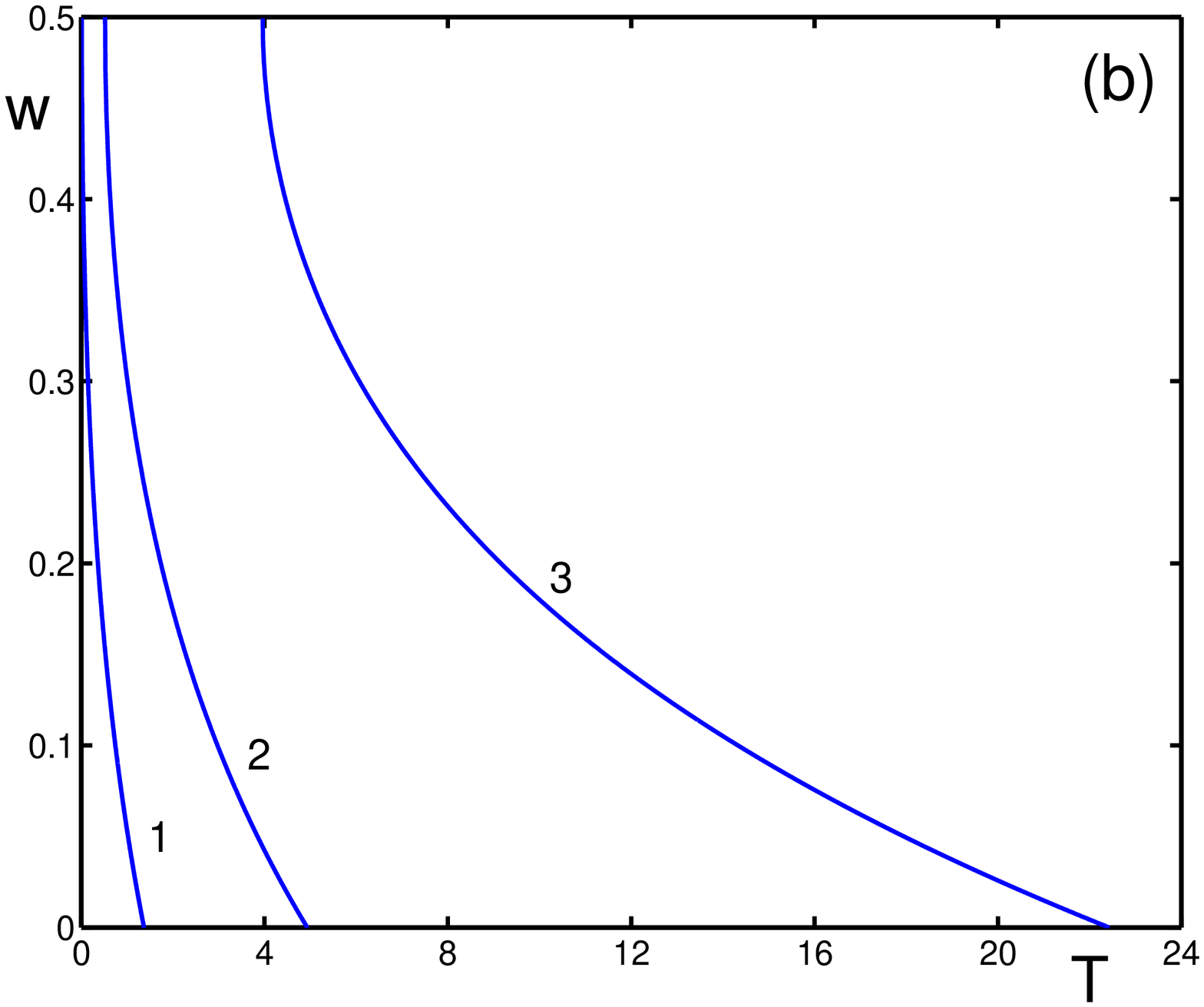}  } }
\caption{
Geometric weight (geometric probability) of the ordered phase $w(T)$ as
a function of dimensionless temperature for different lattice filling 
factors:
(a) $\nu=0.1$ (line 1),  $\nu=0.2$ (line 2), $\nu=0.3$ (line 3), and
$\nu=0.329$ (line 4);
(b) $\nu=0.33$ (line 1),  $\nu=0.4$ (line 2), and $\nu=0.45$ (line 3).
}
\label{fig:Fig.3}
\end{figure}


\newpage

\begin{thebibliography}{99}

\bibitem{Schulze_1}
G.E.R. Schulze,
Metallophysics, Academic, Berlin (1967). 

\bibitem{Ziman_2}
J.M. Ziman, 
Models of Disorder, Cambridge University, Cambridge, 1979.

\bibitem{Filshtinsky_3}
E.I. Grigolyuk, L.A. Filshtinsky,
Regular Piecewise Homogeneous Structures with Defects, Fizmatgiz, Moscow, 1994.

\bibitem{Hirth_4}
J.P. Hirth, J. Lothe, 
Theory of Dislocations, Wiley, New York, 1982.

\bibitem{Hull_5}
D. Hull, D. Bacon,
Introduction to Dislocations, Elsevier, Oxford, 2001.

\bibitem{Gibbs_6}
J.W. Gibbs, 
Collected Works, Longmans, New York, 1928.

\bibitem{Yukalov_7}
V.I. Yukalov,
Phase transitions and heterophase fluctuations,
Phys. Rep. 208 (1991) 395--489.

\bibitem{Yukalov_8}
V.I. Yukalov,
Mesoscopic phase fluctuations: General phenomenon in condensed matter,
Int. J. Mod. Phys. B 17 (2003) 2333--2358. 

\bibitem{Bourbaki_9}
N. Bourbaki, 
Th\'{e}orie des Ensembles, Hermann, Paris, 1958.

\bibitem{Yukalov_10}
V.I. Yukalov,
Systems with symmetry breaking and restoration,
Symmetry 2 (2010) 40--68. 

\bibitem{Kullback_11}
S. Kullback, R.A. Leibler, 
On information and sufficiency,
Ann. Math. Stat. 22 (1951) 79--86. 

\bibitem{Kullback_12}
S. Kullback, 
Information Theory and Statistics, Wiley, New York, 1959.

\bibitem{Cottrell_14}
A.H. Cottrell, 
Dislocations and Plastic Flow in Crystals, Oxford University, London, 1953.

\bibitem{Friedel_15}
J. Friedel, 
Dislocations, Pergamon, Oxford, 1967.

\bibitem{Hirth_16}
J. Hirth, J. Lothe, 
Theory of Dislocations, McGraw Hill, New York, 1968.

\bibitem{Langer_17}
J.S. Langer, 
Thermal effects in dislocation theory, 
Phys. Rev. E 94 (2016) 063004.

\bibitem{Langer_18}
J.S. Langer, 
Thermodynamic theory of dislocation-enabled plasticity, 
Phys. Rev. E 96 (2017) 053005.

\bibitem{Langer_19}
J.S. Langer, 
Statistical thermodynamics of crystal plasticity, 
J. Stat. Phys. 175 (2019) 531--541.

\bibitem{Yukalov_13}
V.I. Yukalov,
Theory of cold atoms: Basics of quantum statistics,
Laser Phys. 23 (2013) 062001.

\bibitem{Yukalov_20}
V.I. Yukalov,
Theory of melting and crystallization,
Phys. Rev. B 32 (1985) 436--446.

\bibitem{Yukalov_21}
V.I. Yukalov,
Effective Hamiltonians for systems with mixed symmetry,
Physica A 136 (1986) 575--587.

\bibitem{Yukalov_22}
V.I. Yukalov,
Renormalization of quasi-Hamiltonians under heterophase averaging,
Phys. Lett. A 125 (1987) 95--100.

\bibitem{Yukalov_23}
V.I. Yukalov,
Procedure of quasiaveraging for heterophase mixtures,
Physica A 141 (1987) 352--374.

\bibitem{Yukalov_24}
V.I. Yukalov,
Lattice mixtures of fluctuating phases,
Physica A 144 (1987) 369--389.

\bibitem{Yukalov_25}
V.I. Yukalov,
Nonequilibrium representative ensembles for isolated quantum systems,
Phys. Lett. A 375 (2011) 2797--2801.

\bibitem{Yukalov_26}
V.I. Yukalov,
Basics of Bose-Einstein condensation,
Phys. Part. Nucl. 42 (2011) 460--513. 

\bibitem{Huang_51}
K. Huang,
Statistical Mechanics, Wiley, New York, 1963. 

\bibitem{Kubo_52}
R. Kubo,
Statistical Mechanics, North-Holland, Amsterdam, 1965.

\bibitem{Isihara_53}
A. Isihara,
Statistical Physics, Academic, New York, 1971. 

\bibitem{Yukalov_27}
V.I. Yukalov,
Matrix order indices in statistical mechanics,
Physica A 310 (2002) 413--434.

\bibitem{Yukalov_28}
V.I. Yukalov,
Order indices and entanglement production in quantum systems,
Entropy 22 (2020) 565. 

\bibitem{Frenkel_29}
J.I. Frenkel, 
Kinetic Theory of Liquids, Clarendon, Oxford, 1946.

\bibitem{Brout_30}
R. Brout, 
Phase Transitions, Benjamin, New York, 1965.

\bibitem{Kirkwood_31}
J.G. Kirkwood, 
Quantum Statistics and Cooperative Phenomena, Gordon and Breach, New York, 1965.

\bibitem{Bogolubov_32}
N.N. Bogolubov, 
Lectures on Quantum Statistics, Gordon and Breach, New York, 1967, Vol. 1.

\bibitem{Bogolubov_33}
N.N. Bogolubov, 
Lectures on Quantum Statistics, Gordon and Breach, New York, 1970, Vol. 2.

\bibitem{Bogolubov_34}
N.N. Bogolubov, 
Quantum Statistical Mechanics, World Scientific, Singapore, 2015.

\bibitem{Yukalov_35}
V.I. Yukalov,
Statistical theory of heterophase fluctuations,
Physica A 108 (1981) 402--416. 

\bibitem{Yukalov_36}
V.I. Yukalov,
Method of thermodynamic quasiaverages,
Int. J. Mod. Phys. B 5 (1991) 3235--3253.

\bibitem{Ono_37}
S. Ono, S. Kondo, 
Molecular Theory of Surface Tension in Liquids, Springer, Berlin, 1960.

\bibitem{Rusanov_38}
A.I. Rusanov,
Thermodynamics of solid surfaces,
Surf. Sci. Rep. 23 (1996) 173--247.

\bibitem{Rusanov_39}
A.I. Rusanov,
Surface thermodynamics revisited,
Surf. Sci. Rep. 37 (2005) 111--239.

\bibitem{Kjelstrup_40}
S. Kjelstrup, D. Bedeaux, 
Non-Equilibrium Thermodynamics of Heterogeneous Systems,
World Scientific, Singapore, 2008.

\bibitem{Bedeaux_41}
D. Bedeaux, S. Kjelstrup,
Fluid-fluid interfaces of multi-component mixtures in local equilibrium,
Entropy 20 (2018) 250. 

\bibitem{Yukalov_42}
V.I. Yukalov,
Properties of solids with pores and cracks,
Int. J. Mod. Phys. B 3 (1989) 311--326.

\bibitem{Yukalov_43}
V.I. Yukalov,
Properties of crystals with local symmetry breaking,
in: Symmetry and Structural Properties of Condensed Matter,
eds. W. Lulek, B. Lulek, M. Mucha, World Scientific, Singapore, 1991.

\bibitem{Yukalov_44}
V.I. Yukalov,
Chaotic lattice-gas model,
Physica A 213 (1995) 482--499. 

\bibitem{Yukalov_45}
V.I. Yukalov, E.P. Yukalova,
Optical lattice with heterogeneous atomic density,
Laser Phys. 25 (2015) 035501.

\bibitem{Bakai_46}
A.S. Bakai, 
Polycluster Amorphous Solids, Sinteks, Kharkov, 2013. 

\bibitem{Boninsegni_47}
M. Boninsegni, A.B. Kuklov, L. Pollet, N.V. Prokof’ev, B.V. Svistunov, M. Troyer, 
Fate of vacancy-induced supersolidity in $^4$He,
Phys. Rev. Lett. 97 (2006) 080401.

\bibitem{Ma_48}
P.N. Ma, L. Pollet, M. Troyer, F.C. Zhang, 
A classical picture of the role of vacancies and interstitials in helium-4,
J. Low Temp. Phys. 152 (2008) 156--163.

\bibitem{Singh_49}
A.K. Singh, E.S. Penev, B.I. Yakobson,
Vacancy clusters in graphane as quantum dots,
ACS Nano 4 (2010) 3510--3514.

\bibitem{Yukalov_50}
V.I. Yukalov,
Saga of superfluid solids,
Physics 2 (2020) 49--66.

\end{thebibliography}
\end{document}